\documentclass[preprint,showpacs,pra]{revtex4}
\usepackage{amsmath}
\usepackage{graphicx}

\input{tcilatex}

\begin{document}

\title{Total electron yields and stopping power of protons colliding with
NaCl-type insulator surfaces II}
\author{Andrea J. Garc\'{\i}a}
\author{J. E. Miraglia}
\date{\today }

\begin{abstract}
We re-calculate electron yields and stopping power of protons colliding with
surfaces of NaCl-type insulators. In this article, the projectile is
considered to move taking into account the static and polarization
potentials with all the individual ions forming the surface lattice unlike
our previous work (Phys. Rev. A\textbf{\ 75}, 042904 (2007)) where the
projectile was considered to move in an homogeneous planar potential.
Substantial differences (up to forty percent of increment) have been found
especially when the projectile incident angle approaches the critical one.
We compare our prediction for electron yield and stopping power with the
available experimental data for LiF, KI and KCl.
\end{abstract}

\affiliation{Instituto de Astronom\'{\i}a y F\'{\i}sica del Espacio. Consejo Nacional de
Investigaciones Cient\'{\i}ficas y T\'{e}cnicas}
\affiliation{Departamento de F\'{\i}sica. Facultad de Ciencias Exactas y Naturales.
Universidad de Buenos Aires. \\
Casilla de Correo 67, Sucursal 28, (C1428EGA) Buenos Aires, Argentina.}
\maketitle

\section{INTRODUCTION}

In our previous article \cite{garcia07} (here refereed to as I), we reported
electron yields and stopping power of proton colliding with surfaces of
sixteen NaCl-type insulators formed with four alkalis Li$^{+}$, Na$^{+}$, K$%
^{+}$ and Rb$^{+}$ and four halides F$^{-}$, Cl$^{-}$ , Br$^{-}$ and I$^{-}$%
. Proton energies were considered to range between 100 kev to 1 Mev. The
(classical) movement of the projectile was calculated under \textit{planar}
channeling condition. In that case, the interaction with the surface is
described by a potential that depends only on the distance to the surface.
In this work, we consider a more realistic surface,\ allowing the projectile
to interact with all the individual ions. In Ref. \cite{garcia06} (hereafter
referred to as II), we called this feature \textit{punctual} channeling .\
We decided to undertake this task -which consumes a long computing time-
because we found substantial differences as compared with the planar model,
due to different trajectories.

The basic considerations we assume here are:

1i) The insulator surface is considered to be composed by an array of alkali
and halide ions at the places given by the crystal parameters, and the local
electronic density is described by the Hartree Fock wave functions of the
isolated ions \cite{clementi}. This is that we call the grid of independent
ion model (GII).

2i) The trajectory of the projectile has been calculated classically,
considering the interaction (static and dynamic polarization) with every
single ion of the grid. The static and dynamic potentials used are tabulated
in Table I of Ref. I

3i) The differential probabilities have been calculated using the Levine
Louie response function \cite{levine} where the gap, in accordance with the
GII model, is considered to be the ionization energy of the isolated ion, as
in I. For each trajectory, the stopping power and the electron yield are
calculated at each segment of the collision and integrated along all the
classical projectile path.

4i) In accordance with the grid of independent ions used, electrons were
counted when the energy provided by the projectile was\ equal or larger than
the binding energy. Following the collision, we assume that there is Auger
decay and so the emission of the inner layers has been multiply by two.

We report here a complete set of electron yields (total number of emitted
electrons) and stopping power of protons colliding grazingly with sixteen
NaCl-type insulator surfaces built with the four alkalis Li$^{+}$, Na$^{+}$,
K$^{+}$ and Rb$^{+}$ and the four halides F$^{-}$, Cl$^{-}$, Br$^{\text{ }-}$
and I$^{-}$. As in Ref. II, we use a Runge Kutta code to describe the
projectile motion. Thirty two projectile nearest neighbors were considered,
i.e. $4\times 4\times 2$. Hundreds of random couple coordinates were
considered to average the incident projectile initial starting point with
respect to the grid. For a given velocity, we have considered 13 incident
polar angles $\theta _{i}$. For each polar angle $\theta _{i}$, we
considered 800 equally-spaced azimuth angles $(\varphi _{i})$ between\ 0\
and 45 degrees with respect to the index [1,0,0]. For each value of $\theta
_{i}\,$and $\varphi _{i}$, around 10$^{4}$ random trajectories were required
for the calculation to converge, and it took several hours of CPU computing
time. Step-to-step, we add the stopping \ and yield produced in the interval
using the same differential probabilities as in I. The present punctual
channeling calculation takes about two to three orders of magnitude more of
CPU computing time than the planar channeling due to the more complicated
projectile trajectories that arise. We think that this effort is justified
because far more physical details are included which are decisive as the
projectile approaches the surface with inclination approaching the critical
angle of penetration.

In Figs. 1 and 2, we plot the total electron production for the sixteen
insulator surfaces as a function of the proton penetration angle normalized
to the critical one (See Table II of Ref. I). We consider all possible
projectiles trajectories including the ones that penetrate and remains in
the bulk. In Figs. 3 and 4 we plot the total energy loss by the proton. When
compared with similar calculations using the planar model (see Figs 5-8 of
Ref. I),\ the following can be stated:

i) At small incident angles both models seem to produce roughly the same
predictions

(although if we look carefully, punctual channeling is always larger).

ii) As we increase the incident angle, the punctual model produces many more
yields and stopping than the planar. At the critical angle ($\theta \approx
\theta _{c}$) the punctual model produces between 37\ and 40\ \% (KI case)
for the total yields and between \ 23 and 31\% (for KI) for the total
stopping.

Next we proceeded to compare with the available experiments. At this stage
it is important to note that, when we compare our results with the
experimental values:

a) We have accounted for coincidence. That is, we consider the electrons
emitted by protons whose trajectories\ penetrated and then left the
insulator or simply rebounded, regardless of the outgoing angle. The ones
that go below the second layer are not considered.

b) We considered that half the electrons emitted do not leave the insulator
because they are lost to the bulk and not counted by the experiment.

c) \ In the theoretical computing we did not account for secondary electron
emissions, i.e. we did not consider cascades produced by primary electrons
during the movement through the insulator.

d) The surface was considered to be perfect without terraces or any other
unevenness.

\ In Fig. 5, the upper figure shows the energy loss as a function of the
incident energy for the KI insulator when the incident angle is 0.5 deg. as
a function of the projectile energy. The theoretical values are in excellent
agreement with the experimental data \cite{Winter07}. The stopping increases
with the projectile energy. In this case the critical angle for 500 kev
protons is of the order of $0.5$ deg., so at larger impact energies the
number of outgoing projectiles decreases.

In the lower figure we plot the electron emission and the stopping as a
function of the proton angle normalized to the critical angle when the
proton energy is 100 kev for the same insulator. In this case the agreement
is not so good as before and we attributed it to the fact that the impact
energy is not large enough yet. The experimental curve shows a soft maximum,
not present in our model.

In Fig. 6 we plot the electron emission for three incident proton energies,
100, 200 and 300 kev on the LiF insulator. In this case, the agreement
between theoretical and experimental values is good except for the lowest
impact energy. Again this\textit{\ }underestimation may be attributed to the
fact that the model is out of range of validity. The experiments in Figs. 5
and 6 correspond to Winter and collaborators \cite{Winter07}.

In Fig. 7 we show a comparison with\ another set of experiments \cite%
{Kimura98, Kimura03} for the electron yield and stopping for KCl and LiF. In
this case, our lower values display a substantial disagreement with the
experimental data. In the case of stopping we have a difference in
percentage of 23 - 36 \% for the LiF and between 15 and 30 \% for the KCl
insulator. For the electron emission, the difference is still grater, about
70 - 80 \%, for the LiF and KCl, respectively. There seems to be an
incompatibility between the experiments, at 300 kev protons where the
experiments account for about 60 electrons while the other one, at 500 kev
account for twice as many. Our theory cannot reproduce this jump.
Experiments are welcome to test the present model.

\section{ACKNOWLEDGMENTS}

The authors acknowledge financial support from CONICET, UBACyT and ANPCyT of
Argentina.

\begin{figure}[h]
\includegraphics[width=0.75\textwidth]{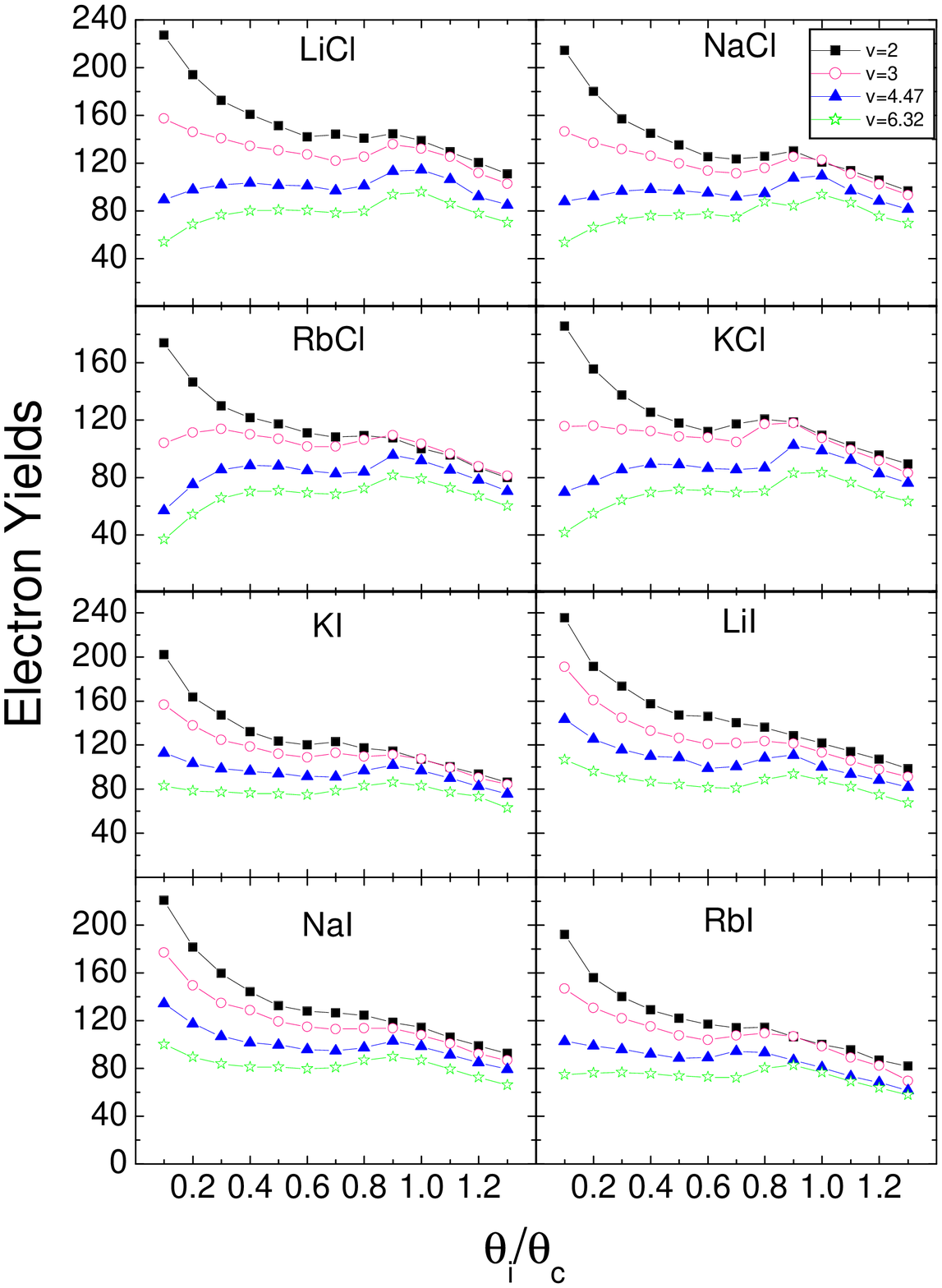}
\caption{Number of total electron production in all directions as a function
of the proton penetration angle normalized to the critical one \protect\cite%
{garcia07} for different velocities and insulators. The calculations
consider all possible projectiles trajectories including the ones that
penetrate and remain in the bulk.}
\label{fig:1}
\end{figure}

\begin{figure}[h]
\includegraphics[width=0.75\textwidth]{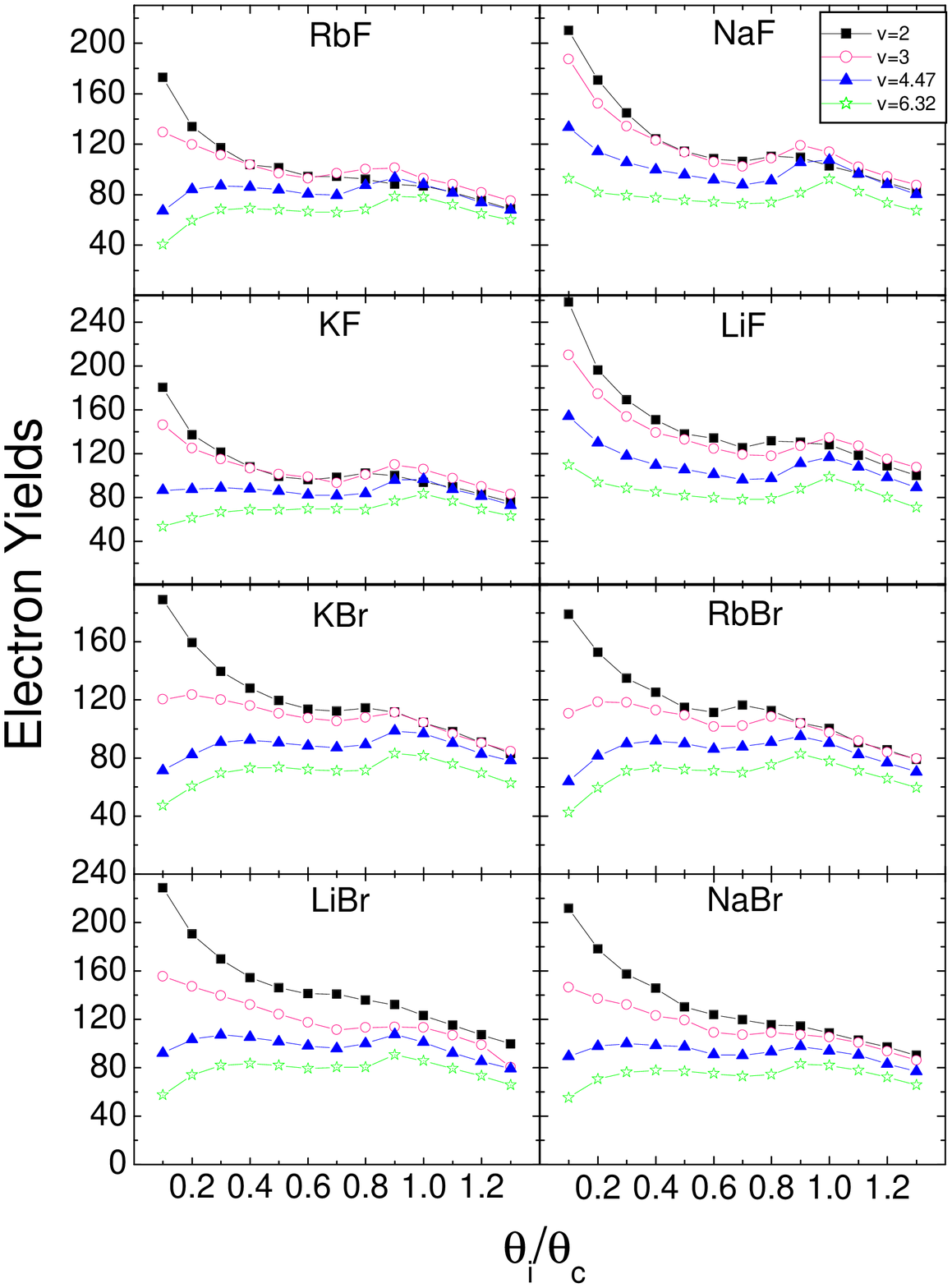}
\caption{Same as Fig.1 for different insulators as indicated.}
\label{fig:2}
\end{figure}

\begin{figure}[h]
\includegraphics[width=0.75\textwidth]{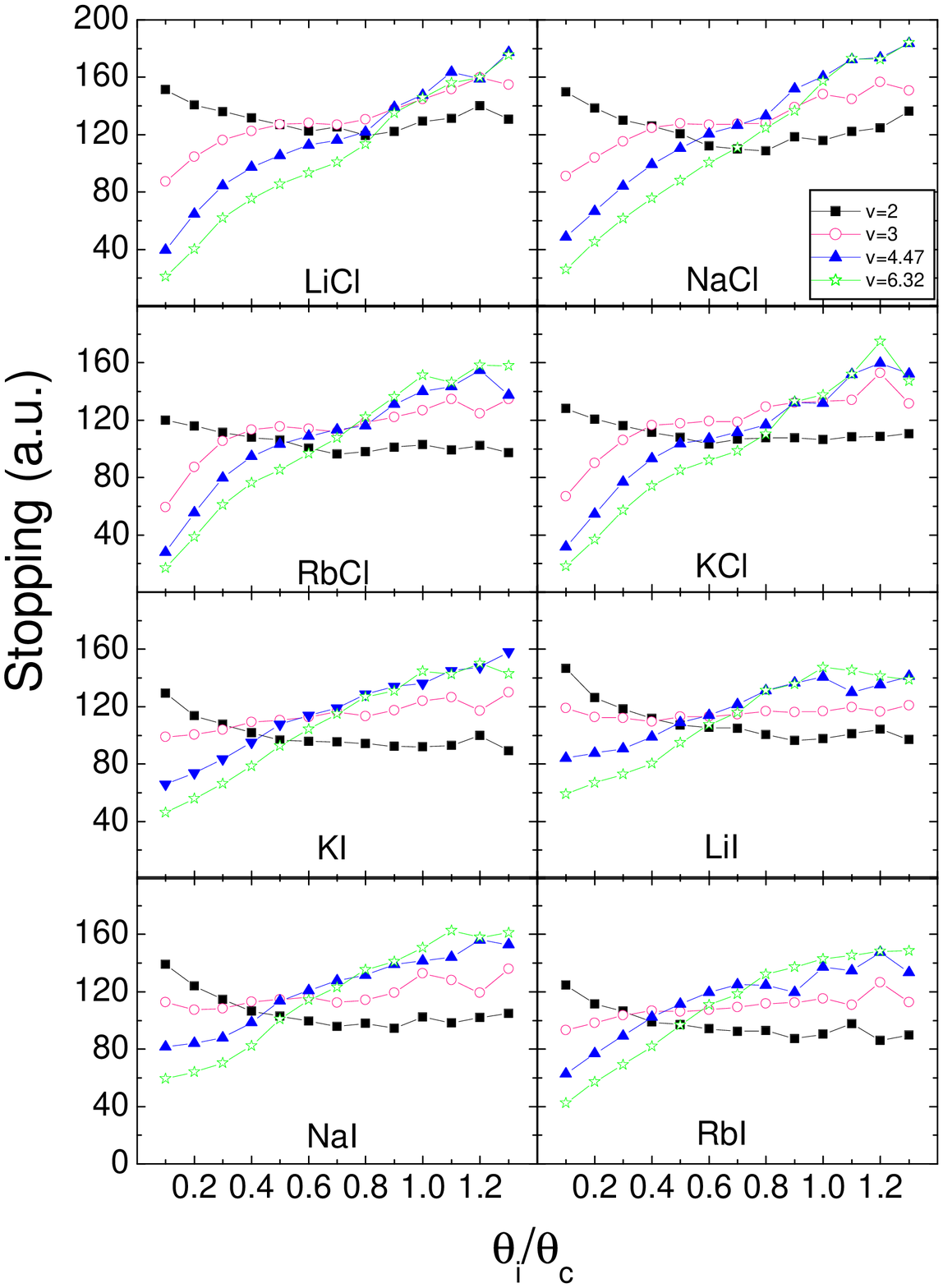}
\caption{Stopping power as a function of the proton penetration angle
normalized to the critical one \protect\cite{garcia07} for different
velocities and insulators.}
\label{fig:3}
\end{figure}

\begin{figure}[h]
\includegraphics[width=0.75\textwidth]{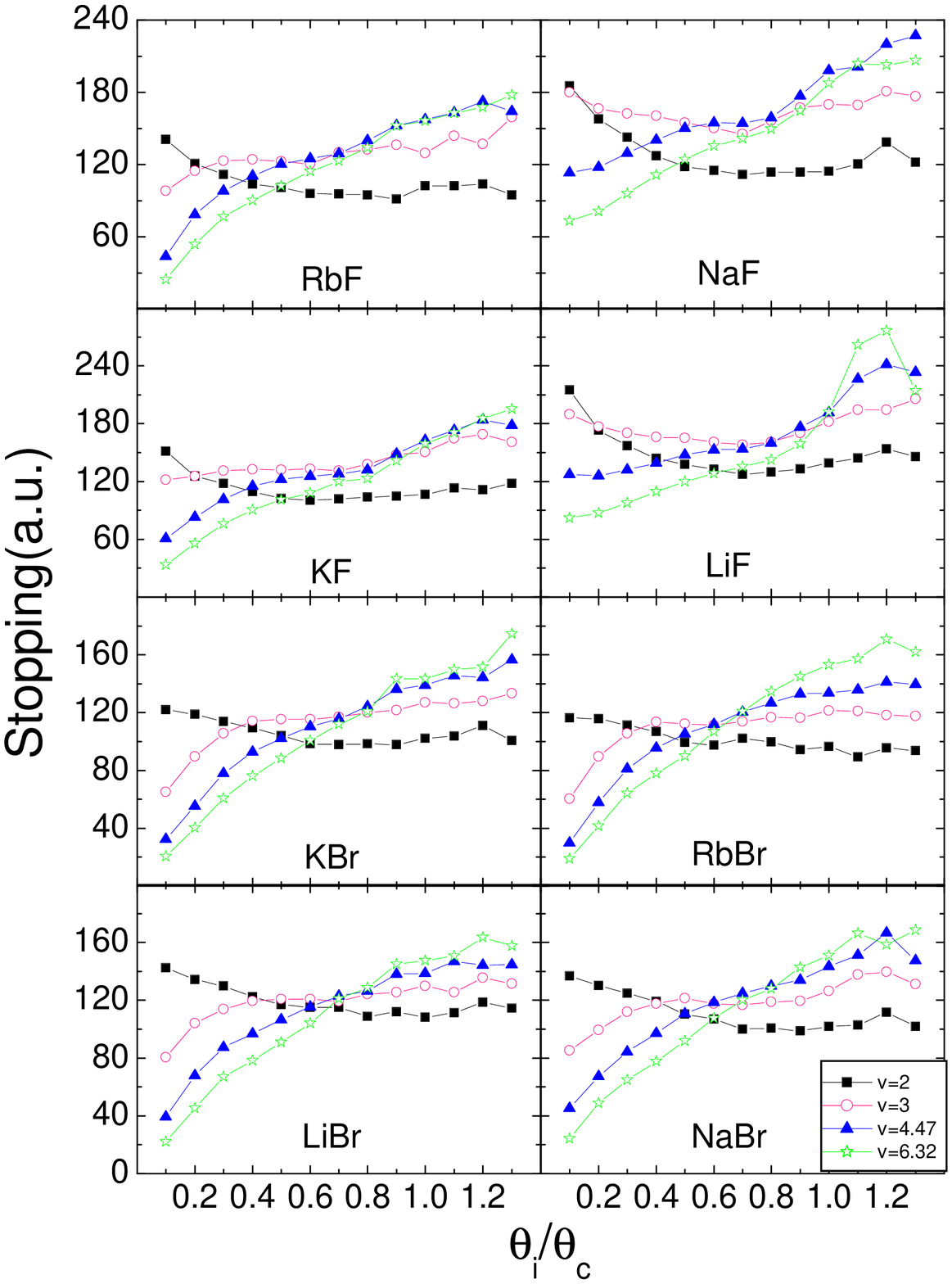}
\caption{Same as Fig.3, for different insulators as indicated.}
\label{fig:4}
\end{figure}

\begin{figure}[h]
\includegraphics[width=0.75\textwidth]{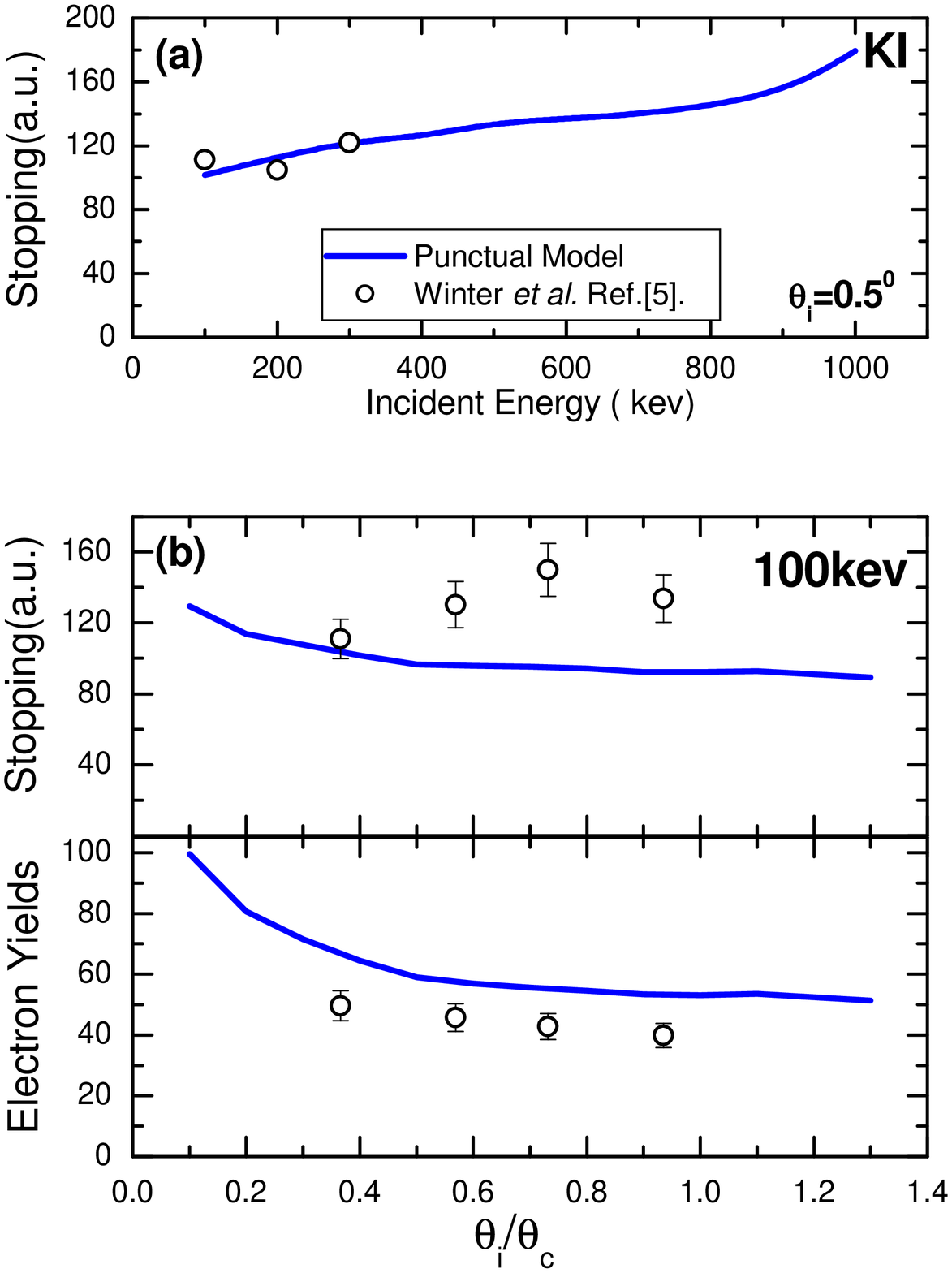}
\caption{(a) Stopping power as a function of the incident energy at incident
angle 0.5 deg. on KI insulator [100] surface. The solid line denotes our
theoretical result and the open circles the experimental values \protect\cite%
{Winter07}. \ (b) Figure above: stopping power ; figure below: electron
emission as a function of the proton penetration angle normalized to the
critical one \protect\cite{garcia07} at 100 kev for KI insulator, calculated
in coincidence when both electron and projectile end in the vacuum. The
solid lines denote the theoretical results and the open circles the
experimental values \protect\cite{Winter07}. }
\label{fig:5}
\end{figure}

\begin{figure}[h]
\includegraphics[width=0.75\textwidth]{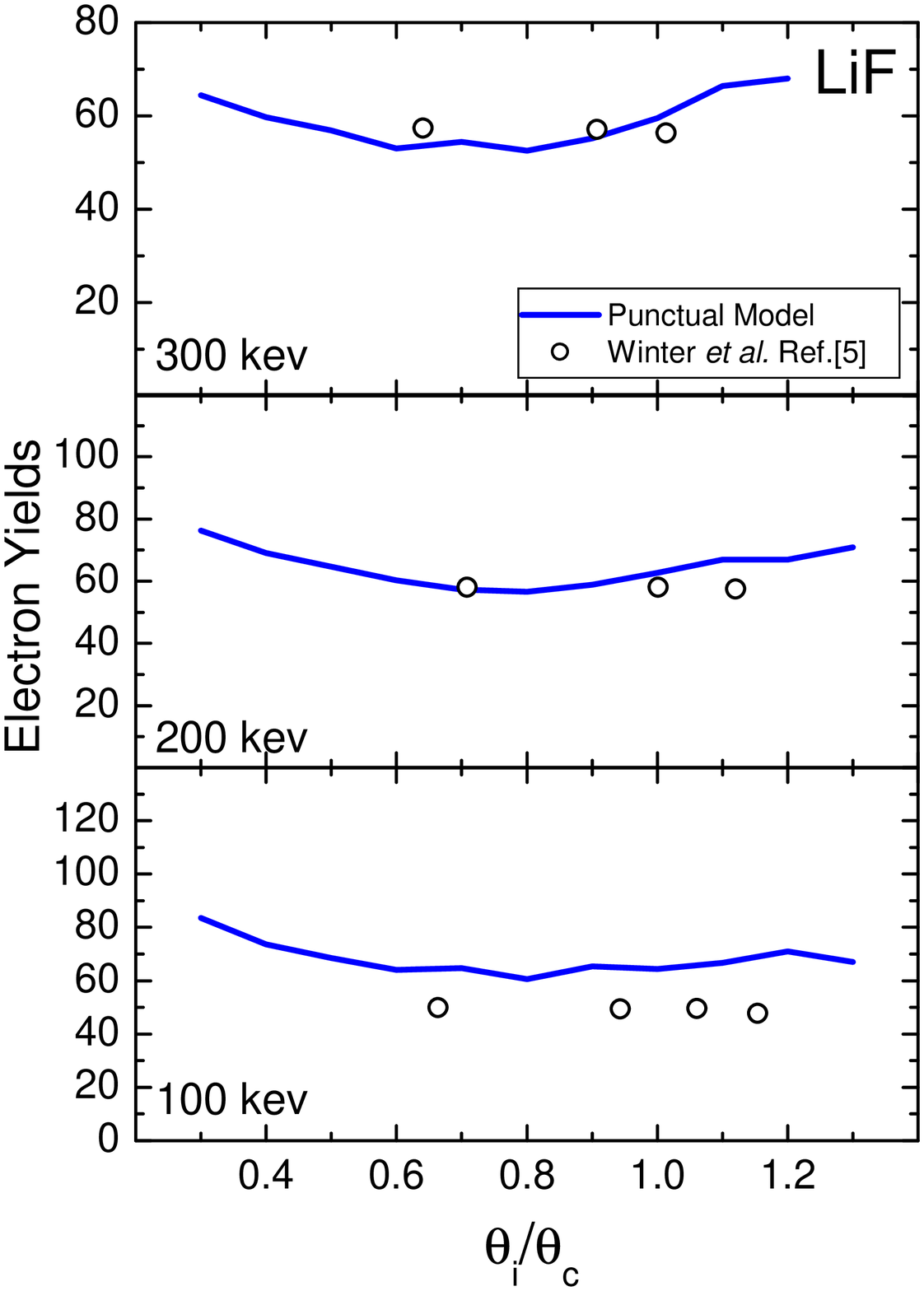}
\caption{Number of total electron production as a function of the proton
penetration angle normalized to the critical one \protect\cite{garcia07} for
different impact energies on LiF insulator [100] surface, calculated in
coincidence when both electron and projectile end in the vacuum. The solid
lines denote the theoretical results and the open circles the experimental
values \protect\cite{Winter07}.}
\end{figure}

\begin{figure}[h]
\includegraphics[width=0.75\textwidth]{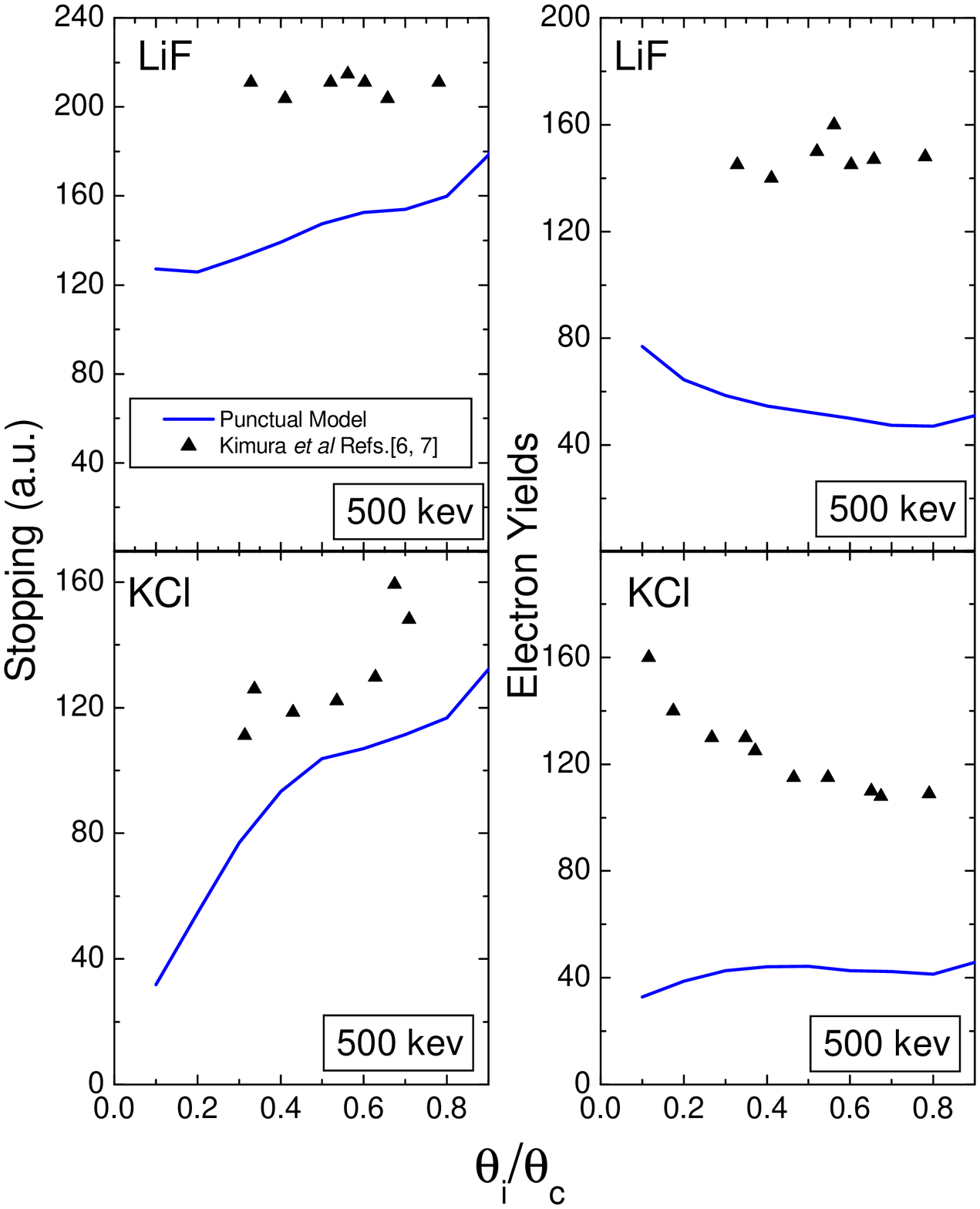}
\caption{Stopping power and electron yields as a function of the incident
energy for KCl and LiF at 500 kev. incident energy The solid lines denote
the theoretical results and the full triangles the experimental values 
\protect\cite{Kimura98, Kimura03}. The electron yields is calculated in
coincidence when both electron and projectile end in the vacuum.}
\label{fig:7}
\end{figure}

\newpage


\begin{thebibliography}{99}
\bibitem{garcia07} A. J. Garc\'{\i}a and J. E. Miraglia, Phys. Rev. A\textbf{%
\ 75}, 042904 (2007).

\bibitem{garcia06} A. J. Garc\'{\i}a and J. E. Miraglia, Phys. Rev. A\textbf{%
\ 74}, 012902 (2006).

\bibitem{clementi} E. Clementi, and C. Roetti, At. Data and Nucl. Data
Tables \textbf{14}, 177 (1974).

\bibitem{levine} Z. H. Levine, S.G. Louie, Phys. Rev. B \textbf{25}, 6310
(1982).

\bibitem{Winter07} M. S. Gravielle , I. Aldazabal, A. Arnau, V. H. Ponce, J.
E. Miraglia, F. Aumayr, S. Lederer and H. Winter, Phys. Rev. A \textbf{76}
(2007) 012904.

\bibitem{Kimura98} K. Kimura, G. Andou, K. Nakajima, Phys. Rev. Lett.\textbf{%
\ 81}, 5438 (1998).

\bibitem{Kimura03} K. Kimura, G. Andou, K. Nakajima, Nucl. Instr. and Meth.
B \textbf{164}, 933 (2000).

\newpage 
\end{thebibliography}
\end{document}